\documentclass[astrosymb,onecolumn,letter]{aastex701}

\newcommand{\HI}{\mbox{H\,{\sc i}}}
\newcommand{\HIbf}{\mbox{H\hspace{0.155 em}{\footnotesize \bf I}}}
\newcommand{\MHI}{$M_{\rm HI}$}

\newcommand{\Msun}{$M_\odot$}
\newcommand{\kms}{\mbox{km\,s$^{-1}$}}
\newcommand{\nan}{Nan\c{c}ay}

\shorttitle{\HI\ in LSBGs}  
\shortauthors{O'Neil, et al. }

\begin{document}

\title{Searching in \HIbf\ for Massive Low Surface Brightness Galaxies: II. Completing the UGC Sample}

\author[0000-0002-2502-5808]{K. O'Neil}
\affiliation{Green Bank Observatory, 155 Observatory Rd, Green Bank, WV 24934, USA}
\email[show]{koneil@nrao.edu}
\author[0000-0003-4770-9829]{W. van Driel}
\affiliation{LUX, Observatoire de Paris, Universit\'e PSL, Sorbonne Universit\'e, CNRS, 
5 place Jules Janssen, 92190 Meudon, France} 
\email[show]{wim.vandriel@obspm.fr}
\author{S. E. Schneider}
\affiliation{University of Massachusetts, Department of Astronomy, 536 LGRC, Amherst, MA 01003, U.S.A.} 
\email[noshow]{redshiftzero@gmail.com}

\begin{abstract} 
Although Low Surface Brightness (LSB) galaxies are increasingly well studied, 
considerable uncertainty still exists as to both their range in properties and their evolutionary history. 
LSB galaxies provide a fascinating extreme in galaxy formation and evolution, allowing 
excellent tests of galaxy formation and evolution theories. Previously, our group successfully 
measured the \HI{} gas and dynamical mass properties of 350 LSB galaxies, significantly 
increasing the number of known extreme LSB galaxies (massive, gas rich, etc.) in the process. 
Here we present 100-m Robert C. Byrd Green Bank Telescope 21-cm \HI{} line observations 
of a further 35 LSB galaxies, only two of which have previously published \HI{} data.
Of the 35, 19 were detected. This paper also includes the \HI{} 
information for five additional galaxies detected accidentally (in the blank sky region
or at alternative velocities), none of which has previously published \HI{} results.
\end{abstract}

      
\section{The Galaxy Sample}  \label{sec:sample}

To explore the number density and properties of LSB galaxies in the nearby Universe, we recently 
undertook a program to re-examine cataloged LSB galaxies without known \HI{} properties 
\citep{2023AJ....165..263O}. A search was made for 21-cm \HI{} line emission in a total of 
350 unique galaxies from two samples whose optical properties indicate they are late-type, LSB galaxies. 
The sample consists of 241 LSB galaxies of morphological type Sb and later selected from the 
HyperLeda database and 119 LSB galaxies from the Uppsala General Catalogue (UGC) with morphological 
types Sd-m and later, selected from the \HI{}-line survey of \citet{schneider90}.
Of the 350 unique galaxies observed, a total of 295 (84\%) were detected in our survey. 

We have continued this successful program, and here we present the results of Green Bank Telescope (GBT) 
\HI{} observations of 35 of the 53 remaining galaxies from our UGC sample. These objects are all 
classified as extreme late types whose disks are low in surface brightness.
All except 5 have previously reported optical redshifts, ranging from 40 to 30,400 \kms. 
This completes the sample of UGC galaxies readily observable by the GBT.

\section{Observations and Data Reduction}  \label{sec:data}

Observations were made with the GBT between August 2024 and March 2026, using the Gregorian 
L-band receiver with the VEGAS spectrometer \citep{2015ursi.confE...4P}. The spectrometer 
was set up to observe from -2000 to 30,250 \kms\ with 0.15 \kms{} velocity resolution, 
which was smoothed to 5 and 15 \kms{} during the data processing phase. 
However, partway through the observations the velocity range in which reliable data could 
be obtained was restricted to v$\leq$ 25,000 \kms, due to Radio Frequency Interference (RFI). 

Position switching was used, with 300s on-source and off-source scans. 
To maximize the scientific output from the telescope, for the majority of the observations
the position-switched observing pattern consisted of one ON+OFF source pair of scans, followed 
by 8 ON source observations, and a final ON+OFF pair, while the remainder were observed using 
the more standard mode with a single ON+OFF pair of scans.

All data were calibrated using the engineering noise diode values measured at the GBT and 
checked by observing a minimum of one standard flux calibrator radio source each observing session.
Data were reduced using standard GBTIDL (http://gbtidl.nrao.edu) routines modified for our 
observing pattern. Frequencies were converted to $cz$ radial velocities in the optical, 
heliocentric velocity frame.

At 21 cm, the GBT half-power beam width is $8\farcm7$$\times$$8\farcm7$.  
A Hubble constant of 70 \kms Mpc$^{-1}$ is used throughout this paper.

\section{Results}  \label{sec:results}
\subsection{Low Surface Brightness Galaxies}  \label{sec:results_LSBs}

All detections of targeted LSB galaxies are listed in Table~\ref{tab:GBT_dets}, together 
with those of five unexpected detections. All values are given using 5 \kms{} resolution.

\begin{itemize}
\item {\bf ID:} Common name for the galaxy;
\item {\bf RA \& Dec:} Right ascension and declination, in J2000.0 coordinates, as used for observations;
\item {\bf Vel$_{HI}$:} Radial \HI{} velocity, heliocentric $cz$ in \kms, defined as the average of the center velocities of the W$_{20}$ and W$_{50}$ line widths;
\item {\bf W$_{20,obs}$, W$_{50,obs}$:} Velocity width at 20\%, respectively 50\%, of the profile's peak height, in \kms, both uncorrected for galaxy inclination;
\item {\bf flux:} Measured integrated \HI{} line flux (Jy \kms);
\item {\bf log(\MHI/\Msun):} Total \HI{} gas mass (\Msun);  
\item {\bf Beam Center:} RA,Dec pointing direction of the GBT for the five unexpected detections; 
\item {\bf Notes:} Notes on GBT observations; details are given at the end of the Table.
\end{itemize}

Two of our target galaxies have previously published \HI{} detections:
NGC 5944 at \nan\ by \citet{2016AA...595A.118V},
and UGC 11977 at Arecibo by \citet{2005ApJS..160..149S} and \citet{2018ApJ...861...49H}.
The published profile parameters are compatible with ours within the estimated uncertainties, 
except for the profile widths of NGC 5944, but neither of the detections is strong.

\subsection{Unexpected Detections}  \label{sec:results_unexpected}

During the data reduction process, five galaxies without previously published \HI{} spectra 
were detected either in the off-source position data or at the on-source observations but 
at a different velocity.  
Of these five, three were readily identified as known sources within the GBT beam, with 
optical velocities which matched our \HI{} values.
However, two other sources do not match known galaxies, and further information on them is 
given below.

{\bf J124113.6-013437:}
None of the many optically identified galaxies within the GBT beam centered on J124113.6-013437 
have a known redshift within less than $\pm$2,500 \kms{} of the detected galaxy.  
Of the remaining 450 galaxies without known velocities, all have been identified by the 
SDSS and WISE surveys \cite{2025arXiv250707093S, 2010AJ....140.1868W}. None of them 
are obvious candidates for our \HI{} detection. 

{\bf J050022.3+305003:}
Of the four known galaxies within the beam (as listed in NED), all have known redshifts 
$>$8,000 \kms{} from the detected galaxy. Even within a 20$^\prime$ radius no obvious 
candidates for this source could be identified.

\section{Discussion}  \label{sec:discussion}

Of the detected galaxies, only one (UGC 11977) can reliably be classified as a massive LSB galaxy, 
with \MHI{} $>$ 10$^{10}$ \Msun. The rest of the galaxies, though, follow an \HI{} mass ?? distribution similar 
to the one we found for the LSB galaxies in \citep{2023AJ....165..263O}.

A comparison of the observed objects with both the ALFALFA galaxy sample \citep{2018ApJ...861...49H} 
with redshifts $z<0.1$ and our previous LSB detections \citep{2023AJ....165..263O} shows 
that overall the sources fit within the distributions. 

Finally, we used the Sloan Digital Sky Survey catalog \citep{2022ApJS..259...35A} to examine 
galaxy colors. Here again, none of the sources appear to be extreme outliers compared to the 
aforementioned samples. Perhaps more importantly, there is no particular distinction between 
the galaxies with and without detected \HI{} gas.

\section{Conclusion}  \label{sec:conclusion}

Here we present the \HI{} properties of the final 35 LSB galaxies from our UGC sample that are 
readily observable by the GBT, and of five other sources detected within the GBT beam whose 
\HI{} properties were previously unknown. 
All in all, the new galaxy sample is unremarkable, adding one new massive LSB galaxy to our 
catalog, and also including four very small dwarf galaxies.
Of the five unexpected detections, three are clearly associated with known spiral galaxies, 
while two do not yet have a clear optical galaxy identification.

Data availability -- The \HI{} spectra supporting Table \ref{tab:GBT_dets} are available via \!\dataset[DOI: 10.5281/zenodo.21536557]{https://doi.org/10.5281/zenodo.21536557}.

\begin{acknowledgements}
This material is based upon work supported by the Green Bank Observatory which is a major 
facility funded by the U.S. National Science Foundation operated by Associated Universities, Inc. 
This research has made use of the NASA/IPAC Extragalactic Database (NED),
the HyperLeda database, SDSS-II data, and POSS-II data.
\end{acknowledgements}

\bibliography{Mybib}{}
\bibliographystyle{aasjournal}

\begin{deluxetable}{lcccccccl}
\tablecaption{All GBT \HI{} detections \label{tab:GBT_dets}}
\tablehead{
\colhead{ID} & \colhead{RA} & \colhead{Dec}   &  \colhead{Vel$_{HI}$}      & \colhead{W$_{20,obs}$} & \colhead{W$_{50,obs}$} & \colhead{flux} & \colhead{log(\MHI/\Msun)} & \colhead{Beam Center}\\
             &  \multicolumn{2}{c}{[J2000.0]} & \colhead{[\kms{}]} & \colhead{[\kms{}]}     & \colhead{[\kms{}]}     & \colhead{[Jy \kms{}]}  & \colhead{} & \colhead{(unexpct. det)}
}
\startdata
\multicolumn{9}{l}{\bf LSB Galaxies}\\
UGC 00134\tablenotemark{*,1}	& 00:14:03.16	&	12:53:53.65	&	1674	$\pm$	5	&	180	$\pm$	5	&	165	$\pm$	5	&	2.13	$\pm$	0.4	&	8.46 & \\
UGC 01134	&	01:35:23.20	&	36:19:06.17	&	5405	$\pm$	5	&	151	$\pm$	5	&	145	$\pm$	5	&	0.61	$\pm$	0.1	&	8.93	&		\\
UGC 01703	&	02:12:55.79	&	32:48:50.81 &	89  	$\pm$	5	&	41	$\pm$	5	&	36	$\pm$	5	&	0.21	$\pm$	0.0	&	4.91	&		\\
UGC 02584	&	03:11:45.04	&	01:02:41.58	&	7607	$\pm$	5	&	290	$\pm$	5	&	281	$\pm$	5	&	1.62	$\pm$	0.3	&	9.65	&		\\
UGC 04979	&	09:22:39.05	&	36:41:13.17	&	7968	$\pm$	5	&	208	$\pm$	5	&	203	$\pm$	5	&	1.36	$\pm$	0.3	&	9.62	&		\\
UGC 05944\tablenotemark{*,2}	&	10:50:18.97	&	13:16:25.51	&	787 	$\pm$	5	&	101	$\pm$	5	&	98	$\pm$	5	&	0.22	$\pm$	0.0	&	6.81	&	\\
UGC 05983\tablenotemark{2*,3}	&	10:52:17.02	&	36:35:43.03	&	613 	$\pm$	5	&	300	$\pm$	5	&	268	$\pm$	5	&	72.75	$\pm$	14.6&	9.12	&	\\
UGC 06395\tablenotemark{*,4}	&	11:22:55.31	&	13:26:31.82	&	4151	$\pm$	5	&	141	$\pm$	5	&	128	$\pm$	5	&	0.70	$\pm$	0.1	&	8.76	&	   \\
UGC 06466	&	11:28:13.16	&	04:18:54.21	&	8244	$\pm$	5	&	233	$\pm$	5	&	230	$\pm$	5	&	0.37	$\pm$	0.1	&	9.08	&		\\
UGC 07553\tablenotemark{*,5}		&	12:27:04.30	&  -01:31:01.86	&	8664	$\pm$	5	&	932	$\pm$	5	&	579	$\pm$	5	&	4.11	$\pm$	0.8	&	10.17	&	\\
UGC 07734	&   12:34:31.55	&	09:37:28.33	&	251 	$\pm$	5	&	38	$\pm$	5	&	36	$\pm$	5	&	0.10	$\pm$	0.0	&	5.48	&		\\
UGC 07929	&	12:45:20.97	&	21:25:36.31	&	393 	$\pm$	5	&	102	$\pm$	5	&	96	$\pm$	5	&	0.28	$\pm$	0.1	&	6.32	&	    \\
UGC 07942\tablenotemark{*,6}	&	12:46:38.43	&	09:18:37.38	&	211 	$\pm$	5	&	45	$\pm$	5	&	41	$\pm$	5	&	0.22	$\pm$	0.0	&	5.67	&	
\\
UGC 07945	&	12:47:01.65	&  -01:34:36.30	&	2739	$\pm$	5	&	218	$\pm$	5	&	182	$\pm$	5	&	2.84	$\pm$	0.6	&	9.01	&		\\
UGC 08123	&	13:00:35.11	&	37:04:17.29	&	4655	$\pm$	5	&	203	$\pm$	5	&	175	$\pm$	5	&	1.34	$\pm$	0.3	&	9.14	&		\\
UGC 08176	&	13:05:14.49	&  -00:22:17.42	&	6780	$\pm$	5	&	382	$\pm$	5	&	352	$\pm$	5	&	1.21	$\pm$	0.2	&	9.43	&		\\
UGC 09551	&	14:50:41.53	&  -01:46:40.66	&	7888    $\pm$	5	&	284	$\pm$	5	&	242	$\pm$	5	&	2.47	$\pm$	0.5	&	9.87	&		\\
UGC 11392	&	19:02:18.71	&	34:47:42.32	&	4487	$\pm$	5	&	215	$\pm$	5	&	197	$\pm$	5	&	1.96	$\pm$	0.4	&	9.28	&		\\
UGC 11977	&	22:17:34.18	&	33:15:27.64	&	11276	$\pm$	5	&	337	$\pm$	5	&	330	$\pm$	5	&	2.50	$\pm$	0.5	&	10.18	&		\\
\hline \\
\multicolumn{9}{l}{\bf Unexpected Detections} \\
UGC 07840	            & 12:41:10.8  & -01:35:26	& 3998  $\pm$ 5 & 210 $\pm$ 5 & 201 $\pm$ 5 & 3.79 $\pm$ 0.4 & 9.46 & 12:41:13.6,-01:34:37 \\
J124113.6-013437*       & --          & --	        & 6860  $\pm$ 5 &  83 $\pm$ 5 & 71  $\pm$ 5 & 0.51 $\pm$ 0.1 & 9.01 & 12:41:13.6,-01:34:37 \\
J050022.3+305003*	    & --          & --	        & 106096$\pm$ 5 
& 383 $\pm$ 5 & 324 $\pm$ 5 & 0.89 $\pm$ 0.2 & 9.68 &  15:00:22.3,30:50:03 \\	
SDSS J095940.16+000239.2&09:59:40     & 00:00:23.9	& 5127  $\pm$ 5 & 172 $\pm$ 5 & 164 $\pm$ 5 & 0.63 $\pm$ 0.1 & 9.00 & 09:59:44.5,00:04:28 \\
VCC 1605	            & 12:35:13.9  & 10:25:53.9  & 1073  $\pm$ 5 & 106 $\pm$ 5 &  94 $\pm$ 5 & 0.05 $\pm$ 0.01& 7.44 & 12:34:59.4,10:28:31 \\	
\enddata
\tablenotetext{*}{Objects marked with an asterisk are considered unreliable detections, as described in these notes. }
\tablenotetext{1}{The detection is clearly of UGC 00132, which lies at 1679 \kms{} and only 4$^\prime$ away \citep{2018ApJ...861...49H}.}
\tablenotetext{2}{This is likely a detection of the galaxy group surrounding NGC 3412.}
\tablenotetext{3}{This likely includes gas from NGC 3432 as well as UGC 5983.}
\tablenotetext{4}{This is likely a detection of IC 2776 or IC 2779, both of which 
lie at $\sim$4155 \kms{} and are 7.6$^\prime$ away from UGC 6395. There is no detection at 
the measured optical velocity of UGC 6395 (852 \kms)}
\tablenotetext{5}{This is approximately 200 \kms{} different from the previously 
measured velocity. That, combined with the wide velocity width and location of two neighbors
10$^\prime$ away (CGCG014-041 \& WISEA J122741.94-012849.2) makes it possible our \HI{} 
profile is confused and contains more than just the \HI{} gas from UGC 07553.}
\tablenotetext{6}{AGC 226177 is an \HI{} source detected by the ALFALFA survey 
which lies 5$^\prime$ away at the same velocity, but without a clear optical counterpart.  
We therefore attribute this detection to UGC 07942.}
\end{deluxetable}

\end{document}